\documentclass[prd,preprint,superscriptaddress,amsmath,amssymb]{revtex4}
\usepackage{graphicx,color,slashed}
\usepackage{graphicx}

\def\be{\begin{eqnarray}}
\def\ed{\end{eqnarray}}
\def\non{\nonumber}

\def\vp{\varepsilon}
\def\ep{\epsilon^*}

\begin{document}

{\begin{flushright}{KIAS-P17017}
\end{flushright}}

\title{\bf \Large Charged-Higgs  on $R_{D^{(*)}}$, $\tau$ polarization, and FBA }

\author{Chuan-Hung Chen}
\email{physchen@mail.ncku.edu.tw}
\affiliation{Department of Physics, National Cheng-Kung University, Tainan 70101, Taiwan}

\author{Takaaki Nomura}
\email{nomura@kias.re.kr}
\affiliation{School of Physics, KIAS, Seoul 130-722, Korea}

\date{\today}

\begin{abstract}
We study the influence of a charged-Higgs on the excess of  branching fraction ratio, $R_M = BR(\bar B \to M \tau \bar\nu_\tau)/BR(\bar B \to M \ell \bar \nu_\ell)$ $(M=D, D^*)$,  in a generic two-Higgs-doublet model. In order to investigate the lepton polarization, the detailed decay amplitudes with lepton helicity are given. When the charged-Higgs is used to resolve  excesses, it is found that two independent Yukawa couplings are needed to  explain the $R_D$ and $R_{D^*}$ anomalies. We show that when the upper limit of $BR(B_c\to \tau \bar \nu_\tau)<30\%$ is included, $R_D$ can be significantly enhanced while $R_{D^*}<0.27$. With the $BR(B_c\to \tau \bar \nu_\tau)$ constraint, we find that the $\tau$-lepton polarizations can be  still affected by the charged-Higgs effects, where the standard model (SM) predictions are obtained as: $P^\tau_{D} \approx 0.324$ and $P^\tau_{D^*}\approx  -0.500$, and they can be enhanced  to be $P^\tau_{D} \approx 0.5$ and $P^\tau_{D^*}\approx  -0.41$ by the charged-Higgs. The integrated lepton froward-backward asymmetry (FBA) is also studied, where the SM result is $\bar A^{D^{(*)},\tau}_{FB}\approx -0.359(0.064)$, and they can be enhanced (decreased) to be $\bar A^{D^{(*)},\tau}_{FB}\approx -0.33 (0.02)$. 
\end{abstract}

\maketitle

\section{Introduction}

The exclusive semileptonic $\bar B \to D^{(*)} \tau \bar\nu_\tau$ decay has drawn a lot of attention since the BaBar collaboration reported evidence for an excess in the ratio of branching ratios (BRs)~\cite{Lees:2012xj,Lees:2013uzd}, which is defined as:
 \begin{equation}
 R_{D^{(*)}} =\frac{BR(\bar B\to D^{(*)} \tau \bar \nu)}{BR(\bar B\to D^{(*)} \ell \bar \nu)}\,.
 \end{equation}
Intriguingly,  a similar excess was also reported by Belle~\cite{Huschle:2015rga,Abdesselam:2016cgx,Hirose:2016wfn} and LHCb~\cite{Aaij:2015yra}.  The averaged results from the heavy flavor averaging group (HFAG) now are given as~\cite{Amhis:2016xyh}:
\begin{align}
R_D &=  0.403 \pm 0.04 \pm 0.024 \,, \nonumber \\
R_{D^*} &=  0.310 \pm 0.015 \pm 0.008\,, \label{eq:RDRDv_data}
\end{align}
where the standard model (SM) predictions are: $R^{\rm SM}_D \approx  0.300$~\cite{Lattice:2015rga,Na:2015kha} and $R^{\rm SM}_{D^*}\approx 0.252$~\cite{Fajfer:2012vx}. Inspired by the unexpected measurements, many implications and speculations were studied in the literature~\cite{Fajfer:2012jt,Crivellin:2012ye,Datta:2012qk,Celis:2012dk,Bailey:2012jg, Sakaki:2013bfa,Ko:2012sv,Crivellin:2015hha, Wei:2017ago,Celis:2016azn,Wang:2016ggf,Freytsis:2015qca,Tanaka:2012nw,Deshpande:2012rr,Dorsner:2013tla,Alonso:2015sja,Bauer:2015knc,Barbieri:2015yvd,Fajfer:2015ycq,Hati:2015awg,Zhu:2016xdg,Deshpand:2016cpw,Becirevic:2016yqi,Sahoo:2016pet,Hiller:2016kry,Das:2016vkr,Li:2016pdv,Calibbi:2015kma,Bhattacharya:2016mcc,Ivanov:2017mrj,He:2012zp,Greljo:2015mma,Boucenna:2016wpr,Boucenna:2016qad,Abada:2013aba,Cvetic:2016fbv,Biancofiore:2013ki,Bhattacharya:2015ida,Alonso:2016gym,Cvetic:2017gkt,Bhattacharya:2016zcw,Li:2016vvp,Alonso:2016oyd,Chen:2017hir,Ivanov:2016qtw,Feruglio:2016gvd,Barbieri:2016las,Bardhan:2016uhr}.

In addition to the branching fraction ratio, Belle recently  showed the measurement of $\tau$ polarization in the $\bar{B} \to D^* \tau \bar\nu_{\tau}$ decay as:
 \begin{equation}
 P^{\tau}_{D^*} = \frac{\Gamma^{h=+}_{D^*} - \Gamma^{h=-}_{D^*}}{\Gamma^{h=+}_{D^*} + \Gamma^{h=-}_{D^*}}=-0.38\pm 0.51^{+0.21}_{-0.16}\,, \label{eq:tauP}
 \end{equation}
 where $\Gamma^{\pm}_{D^*}$ denotes the partial decay rate with a $\tau$-lepton helicity of $h=\pm$, and the SM result is $P^{\tau}_{D^*} \approx -0.500$. 
 Although the errors in the current observation are still large, any significant deviations from the SM results can indicate the new physics~\cite{Tanaka:2012nw,Tanaka:2010se}.  Based on the current experimental indications and  from a phenomenological viewpoint, we investigate the  impacts of a charged-Higgs on the semileptonic $B$ decays in a generic two-Higgs-doublet model (THDM)~\cite{Crivellin:2012ye,HernandezSanchez:2012eg,Benbrik:2015evd}.

Without imposing extra global symmetry, it is known that the flavor-changing neutral currents (FCNCs)  in the THDM can be induced at the tree level. The related couplings can be severely constrained by $\Delta F=2$ processes, such as $K-\bar K$, $B_q-\bar B_q$ (q=d,s), and $D-\bar D$ mixings.  However, these constraints occur in down-type quarks and the first two generations of up-type quarks. The top-quark-involving effects may not have  serious bounds.   Thus, with the constraints from the Higgs precision measurements, the BR for the  $t\to c h$ process  can be of the order of $10^{-4}-10^{-3}$~\cite{Arhrib:2015maa}.  It was found that  the same FCNC coupling in the quark sector can help to resolve $R_D$ and $R_{D^*}$ excesses by the charged-Higgs mediation~\cite{Crivellin:2012ye,HernandezSanchez:2012eg}.  In addition, the tree-level FCNCs in the lepton sector not only can resolve the muon $g-2$ anomaly, but also can provide a large BR to the lepton-flavor violating process, such as $h\to \tau \mu$, for which the details can be seen in our previous work~\cite{Benbrik:2015evd}. 

In this study,  we revisit the charged-Higgs Yukawa couplings to the $\bar B\to (D,D^*)\tau \nu_{\tau}$ processes in the generic THDM. In order to naturally suppress the FCNCs at the tree level, we adopt the so-called Cheng-Sher ansatz~\cite{Cheng:1987rs} in the quark and lepton sectors, where the Yukawa couplings are formulated by $Y_{ij} \propto \sqrt{m_i m_j}/v$. For the semileponic $B$ decays, the main uncertainty is from the hadronic  $\bar B \to D^{(*)}$ transition form  factors. In order to restrain the influence of hadronic effects, we take the values of form factors so that the BRs for  $\bar B\to D^{(*)} \ell \bar\nu_\ell$ ($\ell=e,\mu$) can fit the experimental data within $2\sigma$ errors. Although the charged-Higgs also contributes to the light lepton channels, due to the suppression of $m_\ell/v$, its effects are small to the $\bar B\to D^{(*)} \ell \bar\nu_\ell$ decays. It is found that we need at least two new parameters to explain the $R_{D}$ and $R_{D^*}$ excesses; one is from quark sector, and the other is from the lepton sector. The conclusion is different from that shown in~\cite{Crivellin:2012ye}, in which only one parameter was requested to explain the excesses. However, when the upper limit of $BR(B_c \to \tau \bar\nu_\tau)< 30\%$ obtained in~\cite{Li:2016vvp, Alonso:2016oyd} is taken into account, the allowed parameter space is strictly constrained. It will be seen that $R_{D}$ can be significantly enhanced while the change of $R_{D^*}$ is minor and can be only up to around $0.27$. 

In addition to the ratios of branching fractions, we also investigate the influence of the charged-Higgs on the lepton-helicity asymmetries, and lepton forward-backward asymmetries (FBAs). We find that $\tau$ polarizations in the $\bar B\to (D, D^*) \tau \bar\nu_\tau$ decays are sensitive to the charged-Higgs effects. For the FBA,  the $\bar B\to D^* \tau \bar\nu_\tau$ process is more sensitive to the charged-Higgs effects.

The paper is organized as follows. In Sec.~II, we briefly introduce  the charged Higgs Yukawa couplings to the quarks and leptons in the generic  THDM. We formulate the decay amplitudes with the lepton helicity states, the differential branching ratios, the lepton polarizations, and the lepton FBAs for $\bar B\to (D, D^*) \ell \bar\nu_\ell$ in Sec.~III.  The numerical estimations and discussions are given in Sec.~IV. A summary is shown in Sec.~V.

\section{Charged-Higgs Yukawa Couplings in the THDM}

To obtain the scalar couplings to the quarks and leptons in the  model, the Yukawa sector is written as~\cite{Ahn:2010zza,Benbrik:2015evd}:
\begin{align}
-{\cal L}_Y &= \bar Q_L Y^d_1 D_R H_1 + \bar Q_L Y^{d}_2 D_R H_2 \nonumber \\
&+  \bar Q_L Y^u_1 U_R \tilde H_1 + \bar Q_L Y^{u}_2 U_R
\tilde H_2 \nonumber \\
&+  \bar L Y^\ell_1 \ell_R H_1 + \bar L Y^{\ell}_2 \ell_R H_2 + H.c.\,, 
\label{eq:Yu}
\end{align}
where  the flavor indices are suppressed, $Q^T_L=(u, d)_L$ and $L^T = (\nu, \ell)_L$ are the $SU(2)_L$ quark and lepton doublets, respectively, $Y^f_{1,2}$ are the Yukawa matrices, $\tilde H_2 = i\tau_2 H^*_2$ with $(i\tau_{2})_{11(22)} =0$ and $(i\tau_2)_{12(21)} =1(-1)$, and the Higgs doublets are usually taken as:
\be
H_i &=& \left(
            \begin{array}{c}
              \phi^+_i \\
              (v_i+\phi_i +i \eta_i)/\sqrt{2} \\
            \end{array}
          \right) \label{eq:doublet}
\ed
with $v_i$ being the VEV of $H_i$. Eq.~(\ref{eq:Yu}) can  recover  the type II THDM  if $Y^{d,\ell}_2$ and $Y^{u}_{1}$ vanish.   Since $\phi_1$ and $\phi_2$ are  two CP-even scalars, and they mix, we can introduce a mixing $\alpha$ angle to describe their physical states. The CP-odd and charged scalars are composed  of  $\eta_{1,2}$ and $\phi^\pm_{1,2}$, respectively; therefore,  the mixing angle only depends on the ratio of $v_1$ and $v_2$. Hence, the relations between the physical and weak scalar states are expressed as: 
\be
 h &=& -s_\alpha \phi_1 + c_\alpha  \phi_2 \,, \non \\
 H&=& c_\alpha \phi_1 + s_\alpha \phi_2 \,, \non \\
H^\pm (A) &=& -s_\beta \phi^\pm_1 (\eta_1) + c_\beta \phi^\pm_2 (\eta_2) \,, \label{eq:hHHA}
\ed
where  $h$ is the SM-like Higgs while $H$, $A$, and $H^\pm$ are new particles in the THDM, $c_\alpha (s_\alpha)= \cos\alpha (\sin\alpha)$, $c_\beta  = \cos\beta = v_1/v$, and $s_\beta= \sin\beta = v_2/v$.

 Using Eqs.~(\ref{eq:Yu}) and (\ref{eq:doublet}),    the quark and charged lepton mass matrices can be written as:
\be
{\bf  M}^{f}=  \frac{v}{\sqrt{2}}\left( c_\beta  Y^{f}_1 + s_\beta  Y^{f}_2 \right)\,.
\ed
We can diagonalize ${\bf M}^{f}$ by introducing  unitary matrices $V^f_L$ and $V^f_R$ via  ${\bf m}^f_{\rm dia} = V^f_L {\bf M}^f V^{f \dagger}_R$. Accordingly, the charged Higgs Yukawa couplings  to fermions can be  found as~\cite{Benbrik:2015evd}:
\begin{align}
-{\cal L}^{H^\pm}_Y &= \sqrt{2} \bar d_L {\bf V}^\dagger \left[ - \frac{ \cot\beta}{v } {\bf m_u} + \frac{{\bf X}^u}{ s_\beta}  \right] u_R H^- \nonumber \\ 
&+ \sqrt{2} \bar u_L {\bf V}\left[ - \frac{ \tan\beta}{v } {\bf m_d} + \frac{{\bf X}^d}{  c_\beta}  \right] d_R H^+  \nonumber \\
&+ \sqrt{2} \bar \nu_L  \left[ -\frac{\tan\beta}{v } {\bf m_\ell} + \frac{{\bf X}^\ell }{ c_\beta}  \right] \ell_R H^+ + H.c.\,, \label{eq:YuCH}
\end{align}
where ${\bf V}$ denotes the Cabibbo-Kobayashi-Maskawa (CKM), and ${\bf X^f}$ is defined as:
\begin{equation}
  {\bf X^u}= V^u_L \frac{Y^u_1}{\sqrt{2}}  V^{u\dagger}_R \,, \     {\bf X^d}= V^d_L \frac{Y^d_2}{\sqrt{2}} V^{d\dagger}_R \,, \ 
{\bf X^\ell}= V^\ell_L \frac{Y^\ell_2}{\sqrt{2}} V^{\ell\dagger}_R\,. \label{eq:Xs}
 \end{equation}
 From Eq.~(\ref{eq:YuCH}), it can be seen that $X^{f}_{ij}$   dictates  not only FCNCs but also violation of lepton universality; in addition,  the effects  of $X^{d}$ and $X^{\ell}$ can be further enhanced with a large value of $\tan\beta$, i.e., $c_\beta \ll 1$.  Although the  $Y^{f}_1$ and $Y^{f}_2$ Yukawa matrices basically are  arbitrary free parameters, 
 since  they are related to  the fermion masses,  in order to show the mass-dependence effects, we further adopt the Cheng-Sher ansatz for  $X^{f}_{ij}$ as $X^{f}_{ij} = \sqrt{m^f_i m^f_j}/v\, \chi^{f}_{ij}$~\cite{Cheng:1987rs},  where   $X^{f}_{ij}$ now are suppressed by $\sqrt{m^f_i m^f_j}/v$, and $\chi^{f}_{ij}$  are the free parameters.

 Before we further discuss the effective interactions mediated by the charged Higgs for the $b\to c \ell \bar\nu_\ell$ process, we analyze the effect of ${\bf X}^{u}$ associated with the CKM matrix, such as $({\bf V}^\dagger {\bf X^u})_{bc}$  in Eq.~(\ref{eq:YuCH}). If we assume that ${\bf X^{u}}$ is a diagonal matrix, it can be clearly seen that $({\bf V}^\dagger {\bf X^u})_{bc} = V^*_{cb} X^u_{cc}$. With the Cheng-Sher ansatz, the charged Higgs Yukawa couplings are then suppressed by $m_c/v$. Due to the fact that there is no significant $\tan\beta$ enhancement,  the contribution is expected to be small. However, if $X^{u}_{ij}$ with  $i\neq j$ are allowed,   $({\bf V}^\dagger {\bf X^u})_{bc}$  can be simplified as:
  \begin{align}
  ({\bf V}^\dagger {\bf X^u})_{bc} &= V^*_{ub} X^u_{uc} + V^*_{cb} X^u_{cc} + V^*_{tb} X^u_{tc} \,,  \nonumber \\
 & \approx \frac{\sqrt{m_t m_c}}{v} \chi^u_{tc}\,, \label{eq: VXu}
  \end{align}
 where we have taken $|V_{ub}|< |V_{cb}| \ll V_{tb} \approx 1$ and the Cheng-Sher relation.  With the approximation $\sqrt{m_t m_c}/v \sim V_{cb}$,   it can be seen that $\chi^u_{tc}\sim O(1)$  indicates a large coupling at the $b$-$c$-$H^\pm$ vertex and  could help to resolve the excesses. A similar situation also occurs in the $({\bf V X^d})_{cb}$, which can be expressed as $({\bf V X^d})_{cb}/c_\beta \approx \sqrt{m_b m_s}\chi^{d}_{sb}/v  $.  Since $\chi^d_{sb}$ can be limited by the $\Delta B =2$ process via the mediation of neutral scalar bosons at the tree level, we can ignore its contribution. Hence, we focus on the influence of $\chi^{u}_{tc}$.

  \section{Phenomenological formulations} 
  
Combing the SM and $H^\pm$ contributions, the effective Hamiltonian for $b\to c \ell \bar\nu_{\ell}$ is written as:
 \begin{equation}
 H_{\rm eff}  =\frac{G_F}{\sqrt 2} V_{cb} \left[ (\bar c b)_{V-A}
 (\bar \ell \nu)_{V-A}+ C^{\ell}_L (\bar c b)_{S-P}
 (\bar \ell \nu)_{S-P}+ C^{\ell}_R (\bar c b)_{S+P}
 (\bar \ell \nu)_{S-P}
 \right]\,, \label{eq:Heff}
\end{equation}
where $(\bar f' f)_{V\pm A}=\bar f' \gamma^\mu (1\pm \gamma_5) f$, $(\bar f' f)_{S\pm P}=\bar f' (1\pm \gamma_5) f$,  and the coefficients from the charged Higgs with $g^2/(8 m^2_W) = 1/(2v^2)$ are given by:
\begin{align}
C^{\ell}_L & \approx   - \frac{m_c m_\ell }{m^2_{H^\pm}} \left( 1- \frac{\chi^{\ell}_{\ell \ell} }{s_\beta}\right) \left(1 - \sqrt{\frac{m_t}{m_c}} \frac{ \chi^{u}_{ct} }{c_\beta V_{cb} } \right) \,,  \label{eq:CL} \\
C^{\ell}_R & \approx - \frac{ m_b m_\ell \tan^2\beta }{m^2_{H^\pm}} \left( 1 - \frac{\chi^{\ell}_{\ell \ell} }{s_\beta}\right)\,. \label{eq:CR}
\end{align}
It can be seen that without $\chi^{\ell}_{\ell \ell}$ and $\chi^u_{ct}$, both $C^\ell_{L}$ and $C^\ell_R$ are negative and correspond to the type-II THDM; as a result, they are destructive  contributions to the SM~\cite{Lees:2012xj}. In addition to the sign issue, we also need to adjust $C^\ell_R + C^\ell_L$ and $C^\ell_R - C^\ell_L$ so that $R_D$ and $R_{D^*}$ can be  explained at the same time. In order to demonstrate the effects of the generic THDM, we will discuss the situations with and without $\chi^{\ell}_{\ell\ell}$ and $\chi^u_{ct}$ after we introduce the differential decay rates for the $\bar B\to (D, D^*) \ell \bar \nu_\ell$ decays. 

To calculate the exclusive semileptonic $B$ decays,  we parametrize the $B\to (D, D^*)$ transition form factors as: 
 \begin{align}
  \langle D(p_2)|\bar c \gamma^{\mu} b | \bar B(p_1)\rangle
   &= F_1(q^2)\left[ P^{\mu}-\frac{P\cdot q}{q^2}q^{\mu} \right]  +F_0(q^2)\frac{P\cdot q }{q^2}q^{\mu}, \nonumber \\
\langle D(p_2)| \bar c  b |\overline B(p_1) \rangle & =
({m_B+m_D})
F_S(q^2)\,,
 \end{align}
 \begin{align}
  \langle D^*(p_2,\epsilon)|\overline c\gamma^{\mu}b|\overline B(p_1)\rangle
   &= \frac{V(q^2)}{m_B+m_{D^*}}\epsilon^{\mu\nu\rho\sigma}
     \epsilon^*_{\nu}P_{\rho}q_{\sigma}, \nonumber\\
  \langle D^*(p_2,\epsilon)|\overline c\gamma^{\mu}\gamma_5 b|\overline
  B(p_1)\rangle
   &=2im_{D^*} A_0(q^2)\frac{\epsilon^*\cdot q}{q^2}q^{\mu}
    +i(m_B+m_{D^*})A_1(q^2)\left[\epsilon^*_{\mu}
    -\frac{\epsilon^*\cdot q}{q^2}q^{\mu} \right] \nonumber\\
    &-iA_2(q^2)\frac{\epsilon^*\cdot q}{m_B+m_{D^*}}
     \left[ P^{\mu}-\frac{P\cdot q }{q^2}q^{\mu} \right],
     \nonumber\\
\langle D^*(p_2, \epsilon)| \overline{c} \gamma_5 b |\overline B(p_1)
\rangle & = -i \epsilon^* \cdot q F_P(q^2), \label{eq:ffs}
 \end{align}
where $\epsilon^{0123}=1$, $P=p_1+p_2$, $q=p_1-p_2$; $\epsilon$ is the polarization vector of $D^*$ meson, and $\epsilon \cdot \epsilon^* =-1$. Using
equations of motion $i \slashed{\partial}b=m_b b$ and $\bar c
i\slashed{\partial}=-m_c \bar c $, we obtain the relationships as:
\begin{eqnarray}
F_S(q^2)\approx  \frac{m_B-m_D}{m_b(\mu)-m_c (\mu)} F_0(q^2), \;\;\;
 F_P(q^2)\approx   \frac{2 m_{D^*}}{m_b (\mu)+m_c(\mu)}  A_0(q^2),
\end{eqnarray}
where $m_{b,c}(\mu)$ are the current quark masses at the $\mu$ scale. According to the interactions in Eq.~(\ref{eq:Heff}),  the decay amplitudes for $\bar B\to (D, D^*) \ell \bar\nu_\ell$ are then shown as:
 \begin{align}
 A_{D} &= \frac{G_F}{\sqrt{2}} V_{cb} \left[F_1 \left(P^\mu -\frac{P\cdot q}{q^2}q^\mu \right) (\bar\ell \nu)_{V-A} \right. \non \\
 &+ \left. \left( m_\ell F_0 \frac{P\cdot q}{q^2} + (C^{\ell}_R + C^{\ell}_L)(m_B + m_D) F_S  \right) (\bar\ell \nu)_{S-P}\right]\,, \non \\
A^L_{D^*}&= -i \frac{G_F}{\sqrt{2}} V_{cb} \left\{\ep\cdot q \left( (C^{\ell}_R -C^{\ell}_L) F_P + 2 A_0 \frac{m_{D^*} m_\ell}{q^2}\right) (\bar\ell \nu)_{S-P} \right. \non \\
&+ \left[ (m_B + m_{D^*}) A_1 \left( \ep_\mu(L) - \frac{\ep\cdot q}{q^2} q_\mu\right) - \frac{A_2 \ep\cdot q}{m_B + m_D}  \left(P_\mu -\frac{P\cdot q}{q^2}q_\mu \right)\right] (\bar\ell \nu)_{V-A}\,, \non \\
A^{T}_{D^*} &=  \frac{G_F}{\sqrt{2}} V_{cb} \left[ \frac{V}{m_B + m_{D^*}} \vp_{\mu\nu\rho \sigma} \ep_\nu(T) P^\rho q^\sigma - i (m_B + m_{D^*})A_1 \ep_\mu(T)
\right] (\bar \ell \nu)_{V-A}\,, \label{eq:amps}
 \end{align}
where we have suppressed the $q^2$-dependence in the  form factors, and  $A^L_{D^*}$ and $A^T_{D^*}$ denote the longitudinal and transverse components of $D^*$-meson, respectively. It can be seen  that  the charged Higgs only affects the longitudinal part. 

In order to derive the differential decay rate with a specific lepton helicity,  we set the coordinates of various kinematic variables in the rest frame of $\ell \bar\nu$ invariant mass as:
 \begin{align}
q &=(\sqrt{q^2}, 0 , 0, 0)\,,~ p_{M} = ( E_{M}, 0, 0, p_M)\,,~ E_{M} = \frac{1}{2\sqrt{q^2}}(m^2_B -q^2 - m^2_{M})\,, \\ 
p_{M} &= \frac{\sqrt{\lambda_M}}{2\sqrt{q^2}}\nonumber \,, ~ p_{\nu} =(E_\nu, p_\nu \sin\theta_\ell \cos\phi, p_\nu \sin\theta_\ell \sin\phi, p_\nu \cos\theta_\ell)\,,~p_{\ell} = (E_\ell , - \vec{p}_{\nu})\,, \non \\
\epsilon(L)&=\frac{1}{m_{D^*}}( p_{D^*},0,0, E_{D^*})\,,~\epsilon(\pm) = \frac{1}{\sqrt{2}} (0, \mp 1, - i,0)\,, ~ p_{\ell}=p_{\nu}=\frac{q^2 -m^2_\ell}{2\sqrt{q^2}}\,,
 \end{align}
where $\theta_\ell$ is the polar angle of a neutrino with respect to the moving direction of $M$ meson in the $q^2$ rest frame, and the components of $\vec{p}_{\ell}$ can be obtained from $\vec{p}_{\nu}$ by using  $\pi-\theta_\ell$ and $\phi + \pi$ instead of $\theta_\ell$ and $\phi$. Since the SM neutrino is left-handed, if we neglect its small mass, its helicity can be fixed to be negative; therefore, we focus on the helicity amplitudes of a charged lepton. Accordingly, 
the charged lepton helicity amplitudes for the $\bar B \to D \ell \bar\nu_{\ell}$ decay can be derived  as:
 \begin{align}
 A^{L, h=+}_{D} & = \frac{G_F V_{cb}}{\sqrt{2}} 
 \left( 2 m_\ell \beta_\ell \frac{\sqrt{\lambda_{D}}}{\sqrt{q^2}} F_1 \cos\theta_\ell - 2\beta_\ell \sqrt{q^2} X^{0\ell}_{D}\right)\,, \\
 A^{L, h=-}_{D} & = \frac{G_F V_{cb}}{\sqrt{2}} \left( -2  \beta_{\ell}  \sqrt{\lambda_{D}} F_1 \sin\theta_\ell \right)
\,, \\
 X^{0\ell}_D & =  \frac{m^2_B - m^2_D}{q^2}  m_\ell F_0 + (m_B + m_D) \left( C^{\ell}_R + C^{\ell}_L\right) F_S \,, \label{eq:X0D}
 \end{align}
where $\beta_{\ell} = (1 - m^2_\ell/q^2)^{1/2}$,   $h= +(-)$ denotes the positive (negative) helicity of a charged lepton,  and the detailed spinor states and  derivations of  $(\bar \ell \nu)_{V-A}$ and $ (\bar\ell \nu)_{S-P}$ with polarized leptons are given in the appendix. Although the $D$ meson does not carry spin degrees of freedom,  in order to use similar notation to that in the  $\bar B \to D^* \ell \bar\nu_\ell$ decay, we put an extra index $L$  in $A^{L, h=\pm}_D$. 

With the same approach, we can obtain the helicity amplitudes for the $\bar B\to D^* \ell \bar\nu_\ell$ decay. Since the $D^*$ meson is a vector boson and carries spin degrees of freedom, we separate the lepton helicity amplitudes into longitudinal (L) and transverse (T) parts to show the information for each $D^*$ polarization. Thus, the helicity amplitudes for $\bar B\to D^* \ell \bar\nu_\ell$ with the $D^*$ longitudinal polarization  are found as:
\begin{align}
A^{L,h=+}_{D^*} & = -i\frac{G_F V_{cb}}{\sqrt{2}} \left(2 m_\ell \beta_\ell h^0_{D^*} \cos\theta_\ell - 2 \beta_\ell \frac{\sqrt{\lambda_{D^*}}}{ \sqrt{q^2} } X^{0\ell}_{D^*}\right) \,, \label{eq:ALDv}\\
A^{L,h=-}_{D^*} & =-i\frac{G_F V_{cb}}{\sqrt{2}} \left( -2\sqrt{q^2} \beta_\ell  h^0_{D^*} \sin\theta_\ell \right) \,, \\
 h^0_{D^*}(q^2) &= \frac{ 1}{2 m_{D^*} \sqrt{q^2}}\left[(m_B^2-m_{D^*}^2-q^2)(m_B+m_{D^*})A_1-
\frac{{\lambda_{D^*} }}{m_B+m_{D^*}}A_2\right]\,, \nonumber \\
X^{0\ell}_{D^*} & = m_\ell A_0 + \frac{q^2}{2m_D^* } (C^\ell_R - C^\ell_L) F_P\,.  \label{eq:X0Dv}
\end{align}
It can be seen that the formulae of $A^{L,h=\pm}_{D^*}$ are similar to those of $A^{L, h=\pm}_{D}$. The decay amplitudes with the $D^*$ transverse polarizations are given by:
\begin{align}
A^{T=\pm, h=+}_{D^*} &= i\frac{G_F V_{cb}}{\sqrt{2}}  \left[-\sqrt{2} m_{\ell} \beta_{\ell}  \sin\theta_\ell e^{\mp i \phi} \right]  h^{\pm}_{D^*} \,, \\
A^{T=\pm, h=-}_{D^*} &= \mp i \frac{G_F V_{cb}}{\sqrt{2}}  \left[-\sqrt{2} \sqrt{q^2} \beta_\ell (1\mp \cos\theta_\ell)  e^{\mp i \phi} \right]h^{\pm}_{D^*} \,, \\
h^{\pm}_{D^*} & = \frac{\sqrt{\lambda_{D^*}} }{m_B + m_{D^*}} V \mp (m_B + m_{D^*} ) A_1\,. \nonumber
\end{align}
Since the scalar charged Higgs cannot affect the transverse parts, the   $A^{T=\pm, h=\pm}_{D^*}$  are only from  the SM contributions. According to these helicity amplitudes, it can be clearly seen that due to angular-momentum conservation, the  $A^{L, h=+}_{D}$ and $A^{L(T), h=+}_{D^*}$, which come from $\bar\ell \gamma_\mu (1-\gamma_5)\nu$,  are chiral suppression and proportional to $m_\ell$.  The charged lepton in $\bar\ell (1-\gamma_5) \nu$ prefers the $h=+$ state and in principle has no chiral suppression; however, in our case, chiral suppression exists and  is from  the charged-Higgs Yukawa couplings due to the Cheng-Sher ansatz. 

Including the three-body phase space, the differential decay rates with  lepton helicity and $D^*$ polarization as a function of $q^2$ and $\cos\theta_\ell$ can be obtained as:
\begin{align}
 \frac{d\Gamma^{h=\pm}_{D}}{dq^2 d\cos\theta_\ell} & = \frac{\sqrt{\lambda_{D}}}{512 \pi^3 m^3_B} \beta^2_{\ell}\,  |A^{L,h=\pm }_{D}|^2   \,, \nonumber \\ 
 \frac{d\Gamma^{L(T), h=\pm}_{D^*}}{dq^2 d\cos\theta_\ell}  & = \frac{\sqrt{\lambda_{D^*}}}{512 \pi^3 m^3_B} \beta^2_{\ell}\,  |A^{L(T),h=\pm }_{D^*}|^2  \,. \label{eq:ang_Ga}
\end{align}
The differential decay rates, which integrate out the polar $\theta_\ell$ angle, are shown in the appendix. 
%
In addition to the branching  fraction ratios $R_D$ and $R_{D^*}$, based on Eq.~(\ref{eq:ang_Ga}),  we can also study the lepton helicity asymmetry~\cite{Tanaka:2010se,Tanaka:2012nw,Datta:2012qk} (see also Refs.~\cite{Kalinowski:1990ba, Garisto:1994vz}) and the FBA~\cite{Chen:2005gr,Chen:2006nua}. Helicity asymmetry can be defined as
 \begin{align}
 P^{\ell}_{M} & = \frac{\Gamma^{h= +}_M - \Gamma^{h=-}_M}{ \Gamma^{h= +}_M +  \Gamma^{h=-}_M }\,,
  \end{align}
 where $M=D, D^*$ and  $\Gamma^{h=\pm}_{D^*}=\sum_{\lambda=L,\pm} \Gamma^{\lambda,h=\pm}_{D^*}$ have summed all $D^*$ 
polarizations.  From Eqs.~(\ref{eq:Gah_D}) and (\ref{eq:Gah_D*}), the lepton helicity asymmetries for $\bar B\to (D, D^{*}) \ell \bar\nu_\ell$ with charged Higgs effects  can be found as: 
\begin{align}
 P^{\ell}_{D} & =  \frac{\int^{q^2_{\rm max}}_{m^2_\ell} dq^2 \sqrt{\lambda_D} \beta^4_\ell \left[ \frac{2}{3} \left(m^2_{\ell} -2 q^2 \right) \lambda_D   F^2_1 /q^2 + 2 m^2_\ell q^2 |X^{0\ell}_D|^2\right]}{\int^{q^2_{\rm max}}_{m^2_\ell} dq^2 \sqrt{\lambda_D} \beta^4_\ell \left[ \frac{2}{3}   \left( m^2_{\ell}+ 2 q^2 \right) \lambda_D   F^2_1/q^2 + 2 m^2_\ell q^2 |X^{0\ell}_D|^2\right]}\,,   \\
 P^{\ell}_{D^*} & =  \frac{\int^{q^2_{\rm max}}_{m^2_\ell} dq^2 \sqrt{\lambda_{D^*}} \beta^4_\ell\left[ \frac{2}{3} (m^2_\ell -2 q^2) \left( \sum_{\lambda=L,\pm} |h^{\lambda}_{D^*}|^2\right) + 2(m^2_\ell/q^2) \lambda_{D^*} |X^{0\ell}_{D^*}|^2 \right]}{\int^{q^2_{\rm max}}_{m^2_\ell} dq^2 \sqrt{\lambda_{D^*}} \beta^4_\ell \left[ \frac{2}{3} (m^2_\ell + 2 q^2) \left( \sum_{\lambda=L,\pm} |h^{\lambda}_{D^*}|^2\right) + 2(m^2_\ell/q^2) \lambda_{D^*} |X^{0\ell}_{D^*}|^2 \right]}\,. \label{eq:taup}
 \end{align}

The lepton FBA can be defined as:
\begin{align}
 A^{M,\ell}_{FB} (q^2) &= \frac{\int^{1}_{0} dz (d\Gamma_M/dq^2dz) - \int^{0}_{-1} dz (d\Gamma_M/dq^2dz)}{\int^{1}_{0} dz (d\Gamma_M/dq^2dz) + \int^{0}_{-1} dz (d\Gamma_M/dq^2dz)}\,,
 \end{align}
where $z=\cos\theta_\ell$, and $\Gamma_M$ denotes the total partial decay rate for the $\bar B\to M \ell \bar\nu_\ell$ decay. Accordingly, the FBAs mediated by the  charged Higgs and $W$-boson  in $\bar B \to (D, D^*) \ell \bar \nu_\ell$ are obtained as:
 \be
 A^{D,\ell}_{FB} (q^2)&=& - \frac{2 m_{\ell} \sqrt{\lambda_D} F_1 X^0_D}{H^+_D + H^-_D}\,, \non \\
  A^{D^*, \ell}_{FB} (q^2) &=&  \frac{1}{\sum_{\lambda=L,\pm} (H^{\lambda, +}_{D^*} +H^{\lambda, -}_{D^*})} \left[ -2 m_\ell \frac{\sqrt{\lambda_{D^*}}}{\sqrt{q^2}}
 h^0_{D^*} X^0_{D^*}   +4 q^2 \sqrt {\lambda_{D^*}} A_1 V
\right]. \label{eq:fba}
 \end{eqnarray}
 From Eq.~(\ref{eq:fba}), it can be seen that  the FBAs in the $A^{D,\ell}_{FB}$ and the longitudinal part of $A^{D^*,\ell}_{FB}$  depend on $m_\ell$ and are chiral suppressed.  Due to $m_\tau/m_b \sim 0.4$, which is not highly suppressed, we expect  $\bar B \to D \tau \bar \nu_\tau$ to have a sizable FBA. Moreover, since $A^{D^*,\ell}_{FB}$ does not vanish in the chiral limit, it can  be sizable in a light charged lepton mode.  
 Basically, the observations of the tau polarization and FBA  rely on tau-lepton reconstruction, where the kinematic information is from its decay products; however, since the final state in a tau decay at least involves one invisible neutrino, it is experimentally  challenging to measure the polarization and FBA.  Instead of $\tau$ reconstruction, an approach using the kinematics of visible particles in $\tau$ decays is recently proposed by the authors in~\cite{Alonso:2017ktd}, where  the tau polarization and FBA can be extracted from an angular asymmetry of visible particles in   a tau decay. Accordingly, it is shown that the $\tau \to \pi \nu_\tau$ decay is the most sensitive channel. Based on this new approach,  a statistical precision of $10\%$ can be achieved at  Belle II with an integrated luminosity of 50 ab$^{-1}$. The detailed analysis can be found in~\cite{Alonso:2017ktd}.

\section{ Numerical analysis and discussions}

 \subsection{ Roles of the $\chi^{\ell}_{\tau \tau}$ and $\chi^{u}_{ct}$ parameters}
 
Before presenting  the detailed numerical analysis, we first discuss the influence of $\chi^{\ell}_{\ell \ell}$ and $\chi^{u}_{ct}$ on the $C^{\ell}_{R}$ and $C^\ell_{L}$. From Eqs.~(\ref{eq:X0D}) and (\ref{eq:X0Dv}), it can be seen that the $\bar B\to (D, D^*) \ell \bar \nu_\ell$ decays are  related to $C^\ell_R + C^\ell_L$ and $C^\ell_R - C^\ell_L$, respectively. If the charged-Higgs can resolve the anomalies, the main effects then are on the $\tau \bar\nu_\tau$ modes due to the lepton mass-dependent Yukawa couplings. In the following study, we concentrate on the $\bar B\to (D, D^*) \tau \bar\nu_\tau$ decays. Since the measured $R_{D}$ and $R_{D^*}$ are somewhat higher than the SM predictions, the $H^\pm$ effects should constructively  interfere with the SM contributions.  Thus, we have to require $C^\tau_R \pm C^\tau_L >0$. Furthermore, from Eq.~(\ref{eq:ALDv}),  the charged-Higgs contribution is associated with the $\sqrt{\lambda_{D^*}}$ factor, which represents $|\vec{p}_{D^*}|$ and decreases while $q^2$ increases; as a result, the $H^\pm$ effects on the $BR(\bar B\to D^* \tau \bar\nu_\tau)$ are not as sensitive as those on the $BR(\bar B\to D \tau \bar\nu_\tau)$.  Therefore, to simultaneously enhance $R_D$ and $R_{D^*}$ through the mediation of the charged-Higgs, we conclude $C^{\tau}_R - C^{\tau}_L  >  C^{\tau}_{R} + C^{\tau}_L > 0$; that is, $C^{\tau}_R > 0 > C^\tau_L $.  We note that $C^{\ell}_{R,L}$ from other scalar boson can in general be much larger than the SM contributions, so  the interference effects are not important.  Since the $H^\pm$ couplings are proportional to the $m_\ell \tan\beta/v$, even with a large  $\tan\beta$ case, the effects are naturally limited. Hence, our discussions can be only applied to the charged-Higgs-like case.

Following the analysis above, we now discuss the roles of $\chi^{\ell}_{\tau \tau}$ and $\chi^u_{ct}$ in the $C^\tau_R$ and $C^\tau_L$. From Eq.~(\ref{eq:CR}), it can be seen that without $\chi^{\ell}_{\tau\tau}$, $C^\tau_R < 0$. In other words, we need $\chi^{\ell}_{\tau \tau}$ to tune the sign of $C^\tau_R$ from negative to positive. According to Eq.~(\ref{eq:CL}), if  we take $C^\tau_R >0$  and ignore the $\chi^u_{ct}$ factor, the sign of $C^\tau_L$ is positive, which disfavors our earlier conclusion. Hence, we need  $\chi^u_{ct}$ to flip the positive $C^\tau_L$ to make it negative. In addition to the signs of $C^\tau_R$ and $C^\tau_L$, their magnitudes are also an important issue. The charged-Higgs Yukawa couplings generally depend on the fermion mass, with the exception of  the top-quark, and the couplings are suppressed by the $m_f/v$. To enhance the involved Yukawa couplings, we adopt a scenario with a large value of $\tan\beta$.

{

\subsection{ Numerical analysis of $R_{D^{(*)}}$ and constraint from  $B_c \to \tau \bar \nu_\tau$} 

To estimate the physical quantities of the exclusive semileptonic $B$ decays, we use the $B\to M$ form factors, which  were calculated based on the heavy quark effective theory (HQET) and shown in~\cite{Sakaki:2013bfa,Caprini:1997mu}. The corrections of ${\cal O}(\Lambda_{\rm QCD}/m_{b,c})$ and ${\cal O}(\alpha_s)$ to the form factors can be found in~\cite{Bernlochner:2017jka}. Most phenomena of interest in this work are related to the ratios of squared decay amplitudes, where $V_{cb}$ is cancelled and the influence from the uncertainties of form factors is mild. 
 In order to  fix the parameters in the $B\to M$ transition form factors,  we adopt the parameter values of the form factors so that the BRs for $\bar B \to (D, D^*) \ell \bar \nu_{\ell}$ in the SM satisfy the experimental data within $1\sigma$ errors. 
 With $V_{cb}=3.93\times 10^{-2}$ and the form factors in~\cite{Sakaki:2013bfa,Caprini:1997mu}, we get the BRs for $B^- \to D^{(*)} \ell \bar\nu_\ell$ in the SM as:
 \begin{align}
 BR(B^- \to D \ell \bar\nu_\ell) &\approx  2.26\, [2.16]  \% ~~~\text{HQET [RQM]}\,,  \nonumber \\
 BR(B^- \to D^{*} \ell \bar\nu_\ell) &\approx  5.58\, [5.81] \%  ~~~\text{HQET [RQM]}\,,
 \end{align}
where the experimental data are $BR^{\rm exp}(B^- \to D \ell \bar\nu_\ell)=(2.27 \pm 0.11)\%$ and $BR^{\rm exp}(B^- \to D^{*} \ell \bar\nu_\ell)=(5.69 \pm 0.19)\%$~\cite{PDG}. For comparison, we also show the results  using the form factors calculated in the framework of relativistic quark models (RQM)~\cite{Melikhov:2000yu}.  Using the same values of $V_{cb}$ and form factors, the BRs for $B^- \to (D, D^{*}) \tau \nu_{\tau}$ in the SM can be straightforwardly calculated as:
 \begin{align}
 BR(B^- \to D \tau \bar\nu_\tau )& \approx  0.69\, [0.63]\%  ~~~\text{HQET [RQM]} \,, \nonumber \\
 BR(B^- \to D^{*} \tau \bar\nu_\tau )& \approx   1.43\, [1.42]\% ~~~\text{HQET  [RQM]}\,,
 \end{align}
 where the experimental values are $BR^{\rm exp}(B^- \to D \tau \bar\nu_\tau ) = (0.77 \pm 0.25)\%$ and  $BR^{\rm exp}(B^- \to D^{*} \tau \bar\nu_\tau ) = (1.88 \pm 0.20)\%$~\cite{PDG}. 
   It can be seen that the $\tau \bar\nu_\tau$ measurements  are somewhat larger than the theoretical estimations. 
  The resulted ratios $R_D$ and $R_{D^*}$ and tau polarizations in the SM are given as:
   \begin{align}
   R_D &\approx  0.307,\, \quad R_{D^*} \approx  0.257\,,  \\
   P^{\tau(\mu)}_D & \approx  0.324(-0.962)\,, \quad P^{\tau(\mu)}_{D^*} \approx -0.513 (-0.986)\,.
   \end{align}
 The obtained values of $R_{D,D^*}$ are close to those values shown in~\cite{Lattice:2015rga,Na:2015kha,Fajfer:2012vx}.  We will use the form factors to estimate the $\tau$ polarizations and FBAs. 

  In order to present the charged-Higgs influence on the ratios $R_{D,D^*}$,  we adopt the formulas parametrized as~\cite{Fajfer:2012vx}:
   \begin{align}
   R_D & \approx R^{\rm SM}_D \left[ 1+ 1.5  Re(C^\tau_R + C^\tau_L) +1.0 |C^\tau_R + C^\tau_L|^2 \right] \,, \\
   R_{D^*} & \approx R^{\rm SM}_{D^*} \left[ 1+ 0.12  Re(C^\tau_R - C^\tau_L ) + 0.05 |C^\tau_R - C^\tau_L|^2 \right]\,.
   \end{align}
  Accordingly,  we  show the contours for $R_D$ and $R_{D^*}$ as a function of $\chi^{\ell}_{\tau \tau}$ and $\chi^{u}_{ct}$ in Fig.~\ref{fig:RD_RDv}(a) and of $\tan\beta$ and $\chi^{u}_{ct}$ in Fig.~\ref{fig:RD_RDv}(b), where we  fix $\tan\beta=40$ and $\chi^{\ell}_{\tau\tau} =4$ in the plots, respectively, and  $m_{H^\pm}=400$ GeV in both plots is used.  For clarity,  we use two limits, $\chi^{u}_{ct}=0$ and $\chi^{\ell}_{\tau\tau}=0$, to concretely show their importance in the following discussions. With $\chi^u_{ct}=0$, we obtain $R_{D^*} \sim 0.3$ when $\chi^{\ell}_{\tau\tau} \sim 15$; however, the corresponding value of $R_D$ has been larger than one.  In such case,  the values of $C^\tau_R$ and $C^\tau_L$ are: $C^\tau_R\sim 1.79 \gg C^\tau_L \sim 0$. 
 Since $ B^- \to D \tau \bar \nu_\tau$ is sensitive to $C^\tau_R + C^\tau_L$, when we require $R_{D^*} \sim 0.3$, it can be expected that $R_{D}$ will be significantly enhanced.   With $\chi^{\ell}_{\tau\tau}=0$, we obtain  $R_D \sim 0.35$ with $\chi^u_{ct}\sim 1$, but $R_{D^*} \sim 0.24$, where $C^\tau_{R, L}\sim (-0.13, 0.24)$, which disfavors the earlier conclusion with  $C^\tau_{R}>0$ and $C^\tau_L < 0$. 
  Based on these two limits, it is clear  that  neither $\chi^{u}_{ct}$ nor $\chi^\ell_{\tau\tau}$ can singly resolve the $R_D$ and $R_{D^*}$ anomalies at the same time. From Fig.~\ref{fig:RD_RDv}(a), it can be seen that by properly adjusting both $\chi^{u}_{ct}$ and $\chi^\ell_{\tau\tau}$, the $R_D$ and $R_{D^*}$ excesses can be  explained together. 

\begin{figure}[phtb]
\includegraphics[width=75mm]{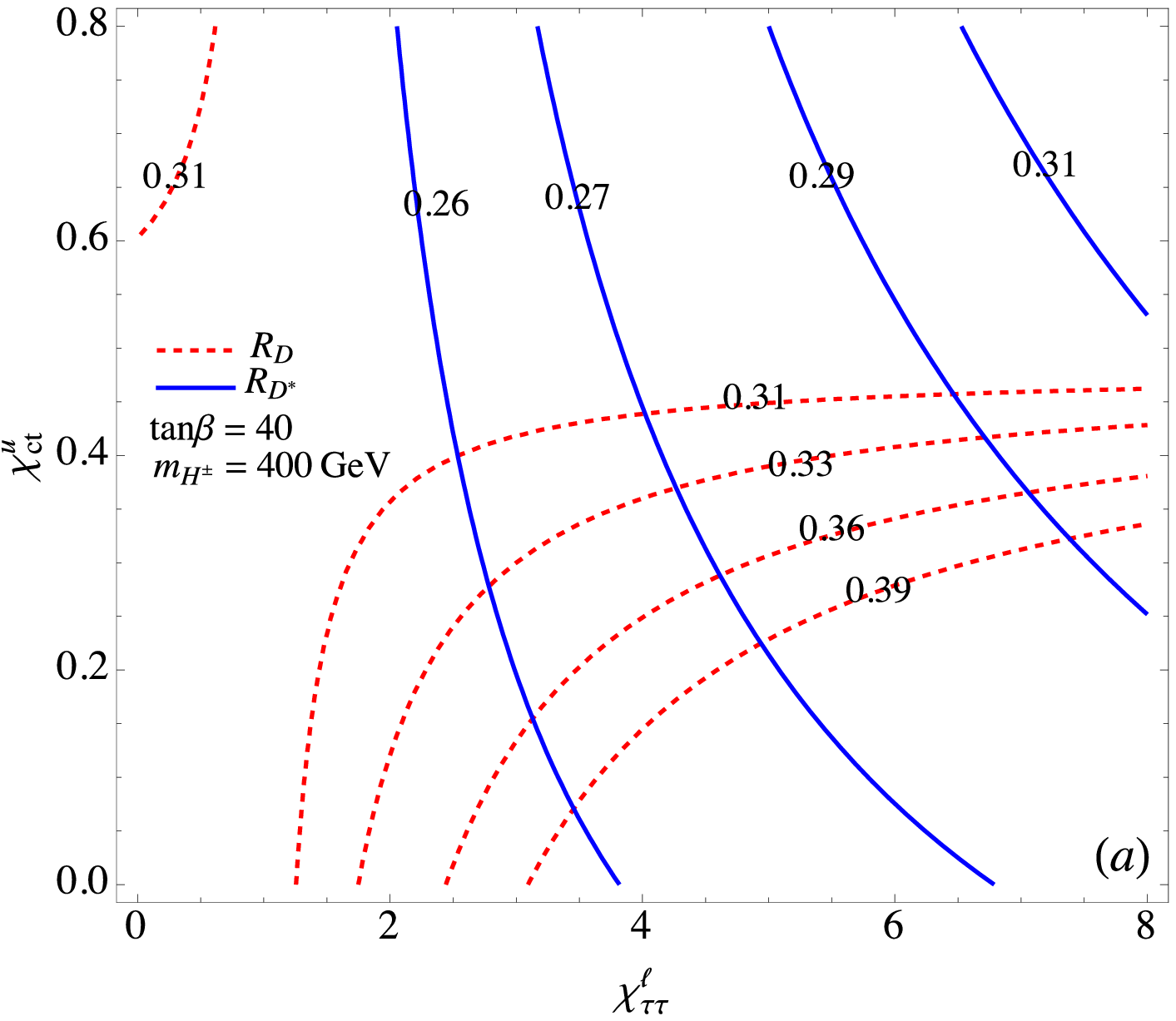}  
\includegraphics[width=75mm]{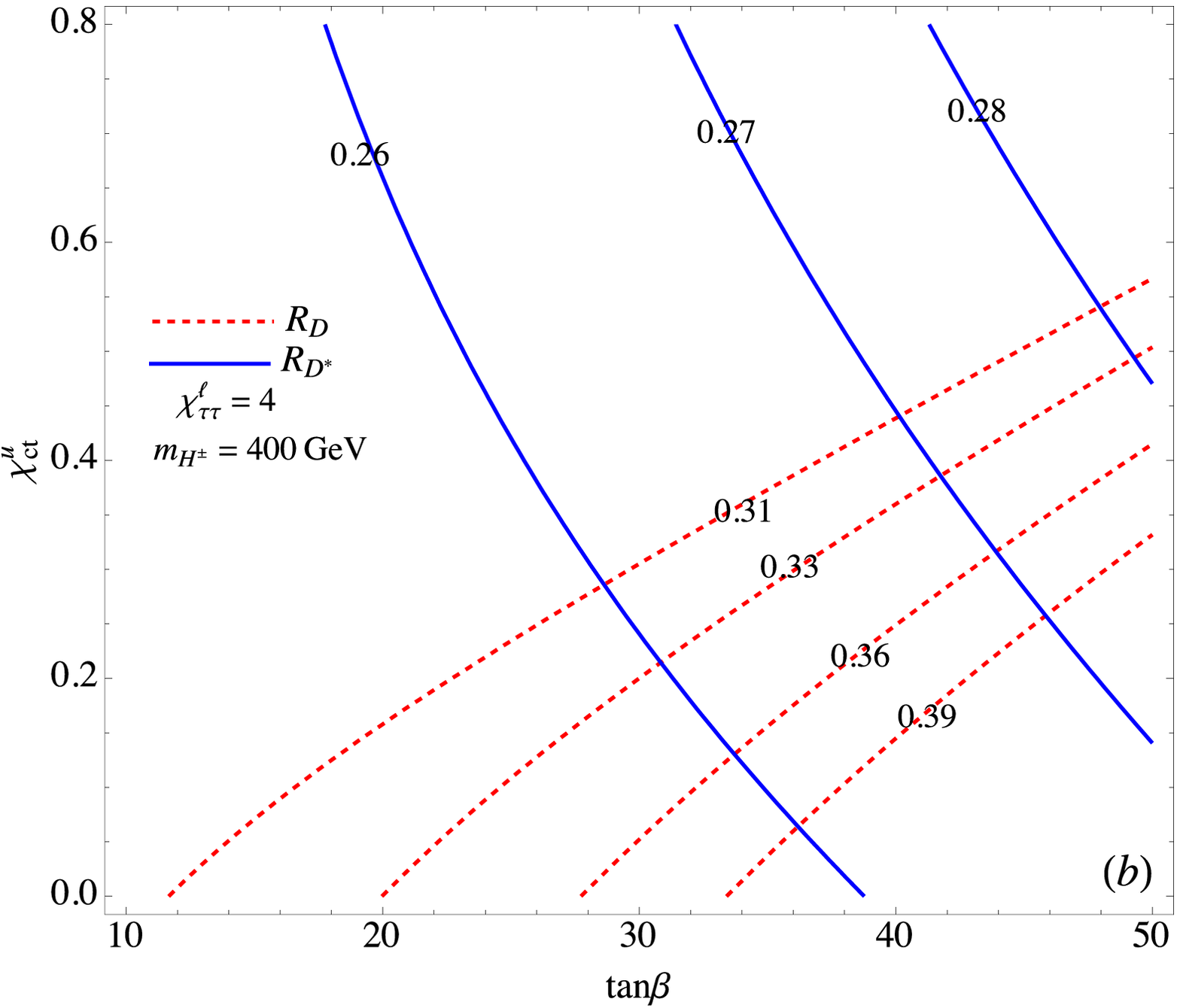} 
\caption{Contours for $R_D$ (dashed) and $R_{D^*}$ (solid) as a function of $\chi^{u}_{ct}$ and (a) $\chi^{\ell}_{\tau \tau}$ with $\tan\beta=40$, (b) $\tan\beta$ with $\chi^{\ell}_{\tau \tau}=4$, where we have fixed $m_{H^\pm}=400$ GeV. }
\label{fig:RD_RDv}
\end{figure}

   In addition to the $B^-\to D^{(*)} \tau \bar\nu_\tau$ decay, the  effective Hamiltonian in Eq.~(\ref{eq:Heff}) also contributes to the $B_c \to \tau \bar\nu_\tau$ process~\cite{Li:2016vvp,Alonso:2016oyd}, where the allowed upper limit, which is obtained from the difference between the SM prediction and the experimental measurement in $B_c$ meson lifetime, is $BR(B_c^- \to \tau \bar\nu_\tau ) < 30 \%$~\cite{Alonso:2016oyd}. We express the BR for $B_c \to \tau \bar\nu_\tau$ as~\cite{Alonso:2016oyd}:
 \begin{align}
 BR(B_c \to \tau \bar \nu_\tau) = \tau_{B_c} \frac{m_{Bc} m^2_\tau f^2_{B_c} G_F^2 |V_{cb}|^2}{8 \pi} \left( 1 - \frac{m_\tau^2}{m_{B_c}^2} \right)^2 
 \left| 1 + \frac{m^2_{B_c}}{m_\tau (m_b + m_c)} \epsilon_P \right|^2,
 \label{eq:BRBc}
 \end{align}
 where $f_{B_c} $ is the $B_c$ decay constant, $\epsilon_{P}=C^\tau_R - C^\tau_L$, and the SM result is  $BR^{\rm SM}(B_c \to \tau \bar \nu_\tau)\approx 2.2\%$. As pointed out by the authors in Refs.~\cite{Li:2016vvp,Alonso:2016oyd}, due to the enhancement factor $m^2_{B_c}/m_\tau (m_b +m_c) \sim 3.6$, the obtained upper limit on $BR(B_c\to \tau \bar \nu_\tau)$ will exclude most of the parameter space for $R_{D^*} > R^{\rm SM}_{D^*}$.  In order to demonstrate the strict constraint, we show the contours for $BR(B_c\to \tau \bar \nu_\tau)$ and $R_{D^{(*)}}$ in Fig.~\ref{fig:RD_RDv_Bc}, where the gray area is excluded by the upper limit of $BR(B_c^- \to \tau \bar\nu_\tau )$. It can be clearly seen that when the constraint from $B_c \to \tau \bar \nu_\tau$ is included, $R_{D}\sim 0.39$ is still allowed; however, $R_{D^*}$ becomes less than 0.28 when $\tan\beta=40$ is used.  Since $R_{D^*}$ is limited by the $B_c \to \tau \bar\nu_\tau$ constraint, hereafter, we just show the range of $R_{D^*}=[0.25,0.27]$.

\begin{figure}[phtb]
\includegraphics[width=75mm]{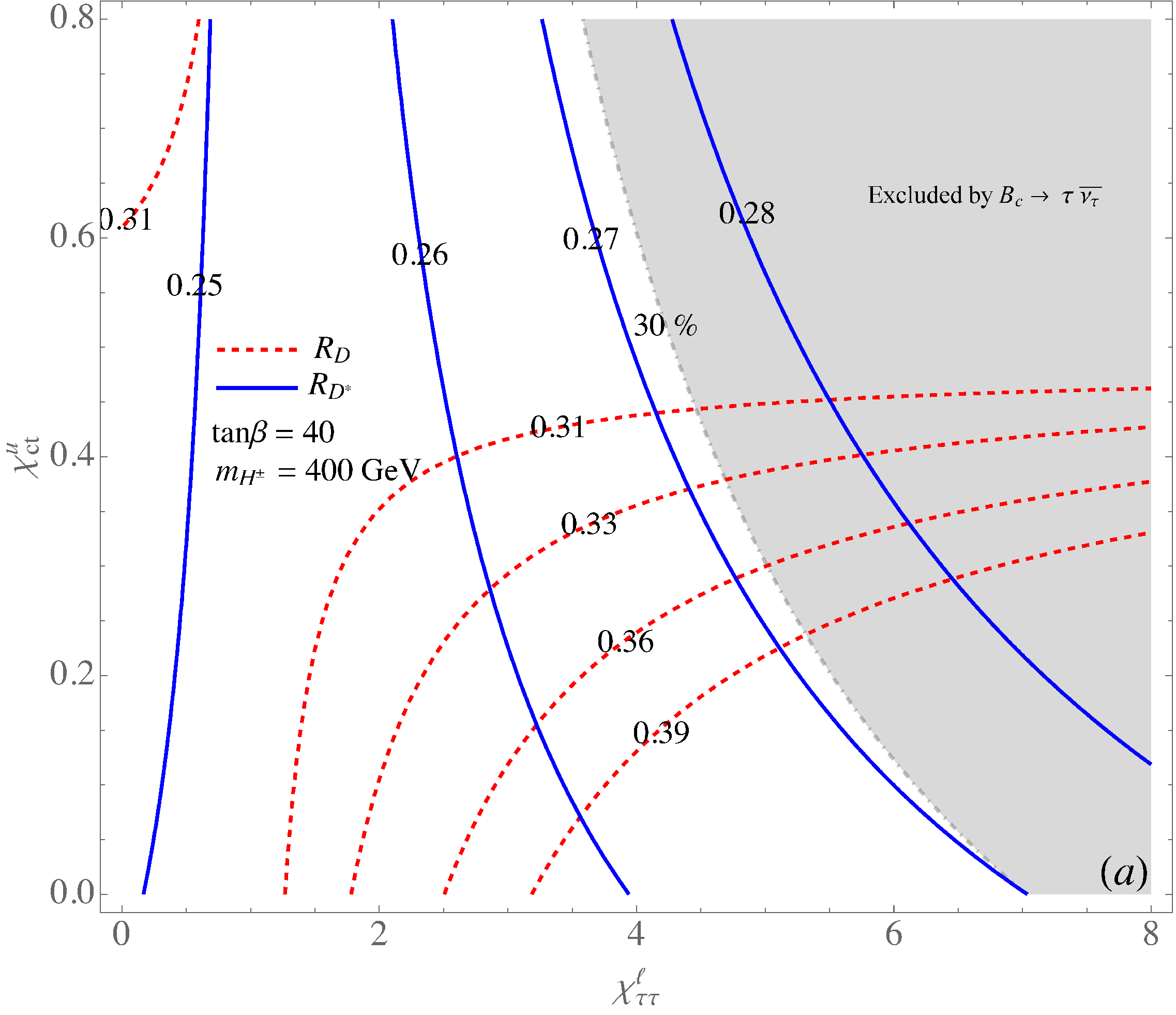}  
\includegraphics[width=75mm]{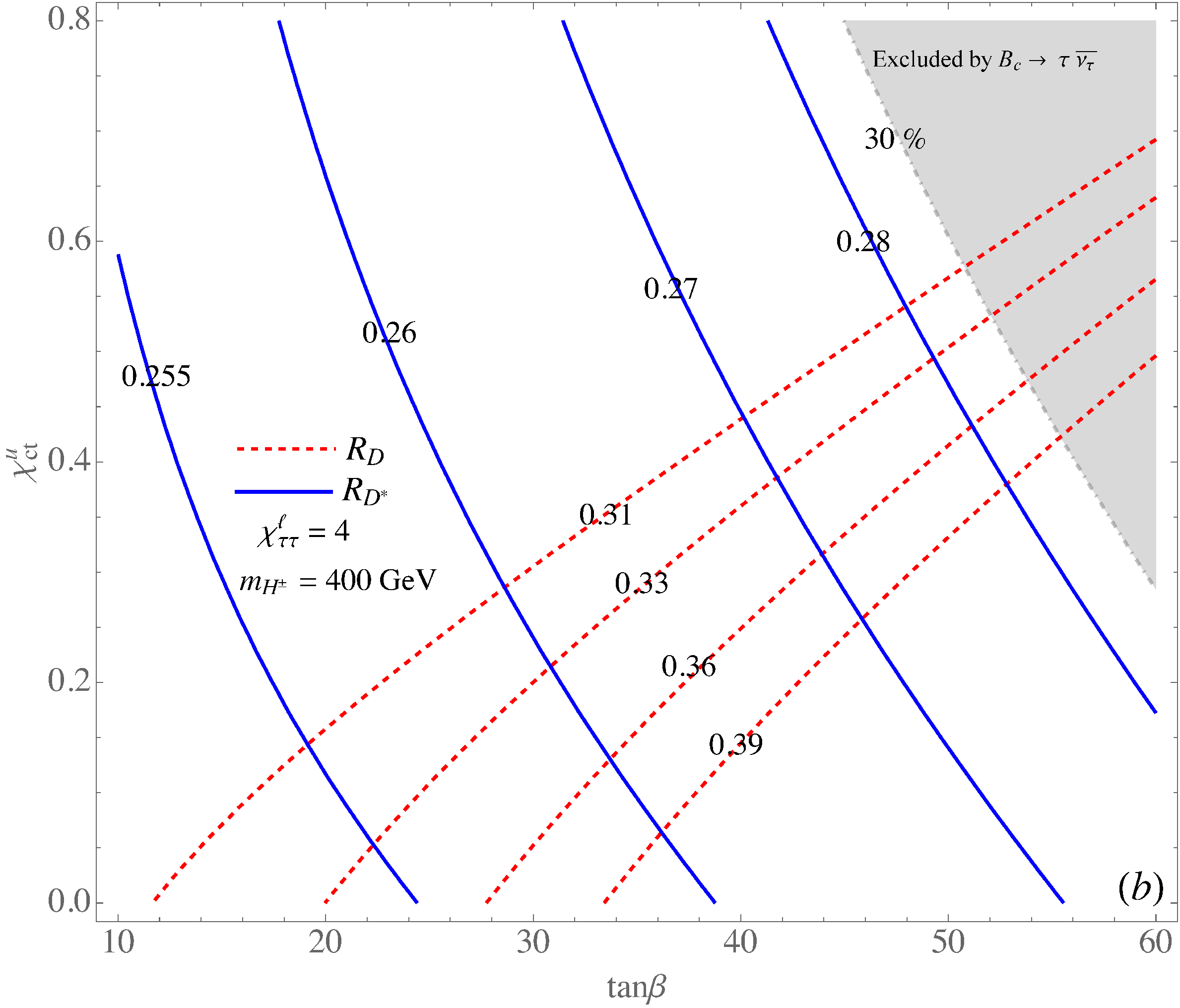} 
\caption{The Legend is the same as that in Fig.~\ref{fig:RD_RDv}, and  the upper limit of $BR(B_c \to \tau \bar\nu_\tau)< 30\%$ is included. }
\label{fig:RD_RDv_Bc}
\end{figure}

\subsection{$\tau$ polarization and FBA}

The Belle collaboration recently reported the measurement of $\tau$ polarization in the $\bar B\to D^* \tau \bar\nu_\tau$ decay. If the $R_D$ and $R_{D^*}$ excesses originate from the new physics, the $\tau$ polarization can be also influenced. To understand the charged-Higgs contributions, according to Eq.~(\ref{eq:taup}), we plot the contours for $P^\tau_{D}$ (dashed) and $P^\tau_{D^*}$ (solid) as a function of $\chi^u_{ct}$ and $\chi^\ell_{\tau \tau}$  with $\tan\beta=40$ in Fig.~\ref{fig:Ptau}(a), and the dependence of $\chi^{u}_{ct}$ and $\tan\beta$ with $\chi^{\ell}_{\tau\tau}=4$ is shown in Fig.~\ref{fig:Ptau}(b), where the bounded areas denote the ranges of $R_{D}=[0.31,0.39]$ (red) and $R_{D^*}=[0.25,0.27]$ (blue).   From the numerical analysis, it can be seen that $P^\tau_{D}$ and $P^\tau_{D^*}$ are more sensitive to $\chi^u_{ct}$ and  $\chi^{\ell}_{\tau\tau}$,  and they can be enhanced from $0.3$ to $0.5$ and $\sim -0.5$ to $\sim -0.41$,  respectively.

\begin{figure}[phtb]
\includegraphics[width=75mm]{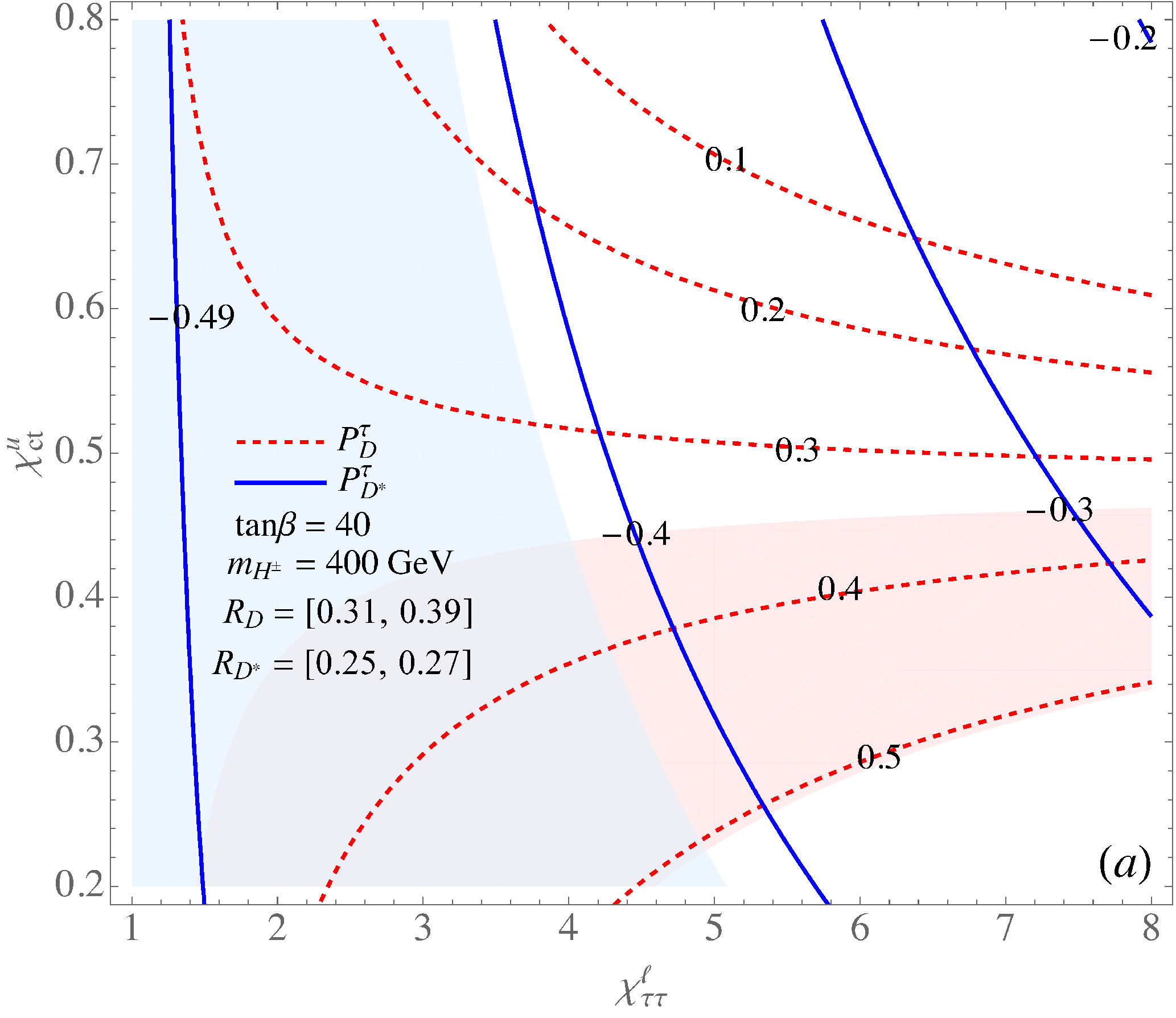}  
\includegraphics[width=75mm]{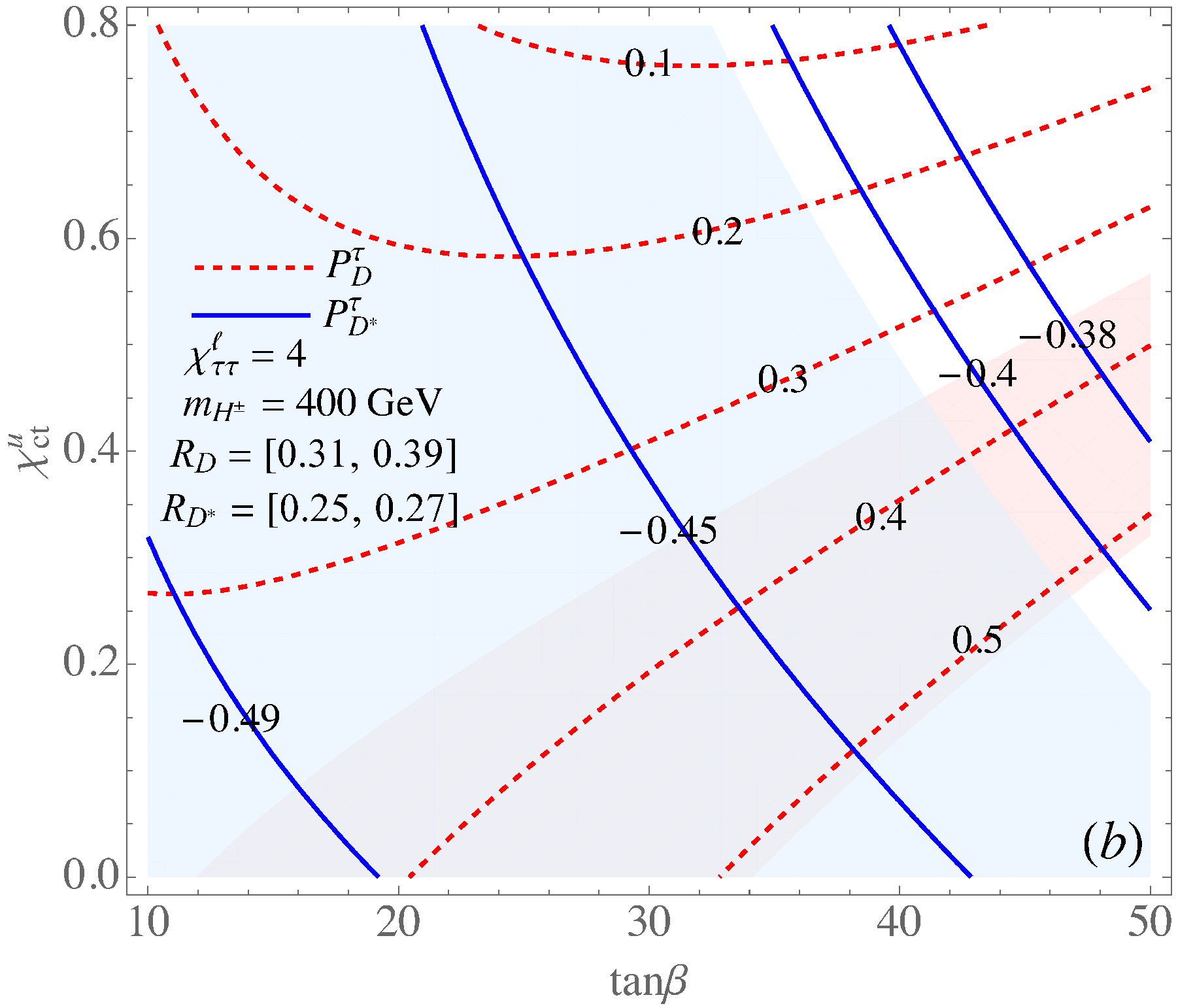} 
\caption{Contours for $P^\tau_D$ (dashed) and $P^\tau_{D^*}$ (solid) as a function of $\chi^{u}_{ct}$ and (a) $\chi^{\ell}_{\tau\tau}$ with $\tan\beta=40$, (b) $\tan\beta$ with $\chi^{\ell}_{\tau\tau}=4$, where we have fixed $m_{H^\pm}=400$ GeV. The bounded areas denote  the ranges of $R_{D}=[0.31,0.39]$ (red) and $R_{D^*}=[0.25,0.27]$ (blue). }
\label{fig:Ptau}
\end{figure}

Finally, we show the numerical results for the FBA.  Since the $\tau$ lepton FBAs have not yet been observed, to demonstrate  the charged-Higgs contributions, we adopt the integrated FBAs, which are defined as:
\begin{align}
 \bar A^{D,\tau}_{FB} &= - \frac{\int^{(m_B -m_D)^2}_{m^2_\tau} dq^2  \left( 2 m^2_{\tau} \lambda_D \beta^4_\tau F_1 X^{0\tau}_D \right)}{ \int^{(m_B -m_D)^2}_{m^2_\tau} dq^2 \sqrt{\lambda_D } \beta^4_\tau (H^+_D + H^-_D)}\,, \non \\
  \bar A^{D^*, \tau}_{FB} &=  \frac{  \int^{(m_B -m_{D^*})^2}_{m^2_\tau} dq^2 \sqrt{\lambda_{D^*} } \beta^4_\tau \left(-2 m_\tau \frac{\sqrt{\lambda_{D^*}}}{\sqrt{q^2}} h^0_{D^*} X^{0\tau}_{D^*}   +4 q^2 \sqrt {\lambda_{D^*}} A_1 V \right)}{\int^{(m_B -m_{D^*})^2}_{m^2_\tau} dq^2 \sqrt{\lambda_{D^*} } \beta^4_\tau \sum_{\lambda=L,\pm} (H^{\lambda, +}_{D^*} +H^{\lambda, -}_{D^*})} \,. \label{eq:Ifba}
 \end{align}
 The contours for $\bar A^{D,\tau}_{FB}$ and $\bar A^{D^*,\tau}_{FB}$ as a function of $\chi^{u}_{ct}$ and  $\chi^{\ell}_{\tau\tau}$ with $\tan\beta=40$ are shown in Fig.~\ref{fig:AFB}(a), whereas the contours with $\chi^{\ell}_{\tau\tau}=4$ as a function of  $\chi^{u}_{ct}$ and  $\tan\beta$ are given in Fig.~\ref{fig:AFB}(b). 
From the plots, it can be seen that $\bar A^{D^*,\tau}_{FB}$ is more sensitive to the charged Higgs effects.  The sign of $\bar A^{D^*,\tau}_{FB}$  in general can be changed in different parameter space; however, when the $BR(B_c\to \tau \bar\nu_\tau)$ constraint is included, its sign  can be only positive.  Moreover,  the magnitude of $\bar A^{D^*,\tau}_{FB}$ is smaller than that of $\bar A^{D,\tau}_{FB}$. These behaviors can be  understood from Eq.~(\ref{eq:Ifba}), in which  the first and second terms in the numerator are opposite in sign and are from the $D^*$ longitudinal and transverse parts, respectively. Although the first term is proportional to $m_\ell$, when $m_\ell = m_\tau$, the contribution from the first  becomes compatible with that  of the second so that the sign can be flipped.   The results with some chosen benchmarks of $(\chi^u_{ct}, \chi^\ell_{\ell \ell})$ are given in Table~\ref{tab:FBA}. For comparisons, we also show the values of  $R_D$, $R_{D^*}$, $P^\tau_D$, and $P^\tau_{D^*}$ in the table.

\begin{figure}[phtb]
\includegraphics[width=75mm]{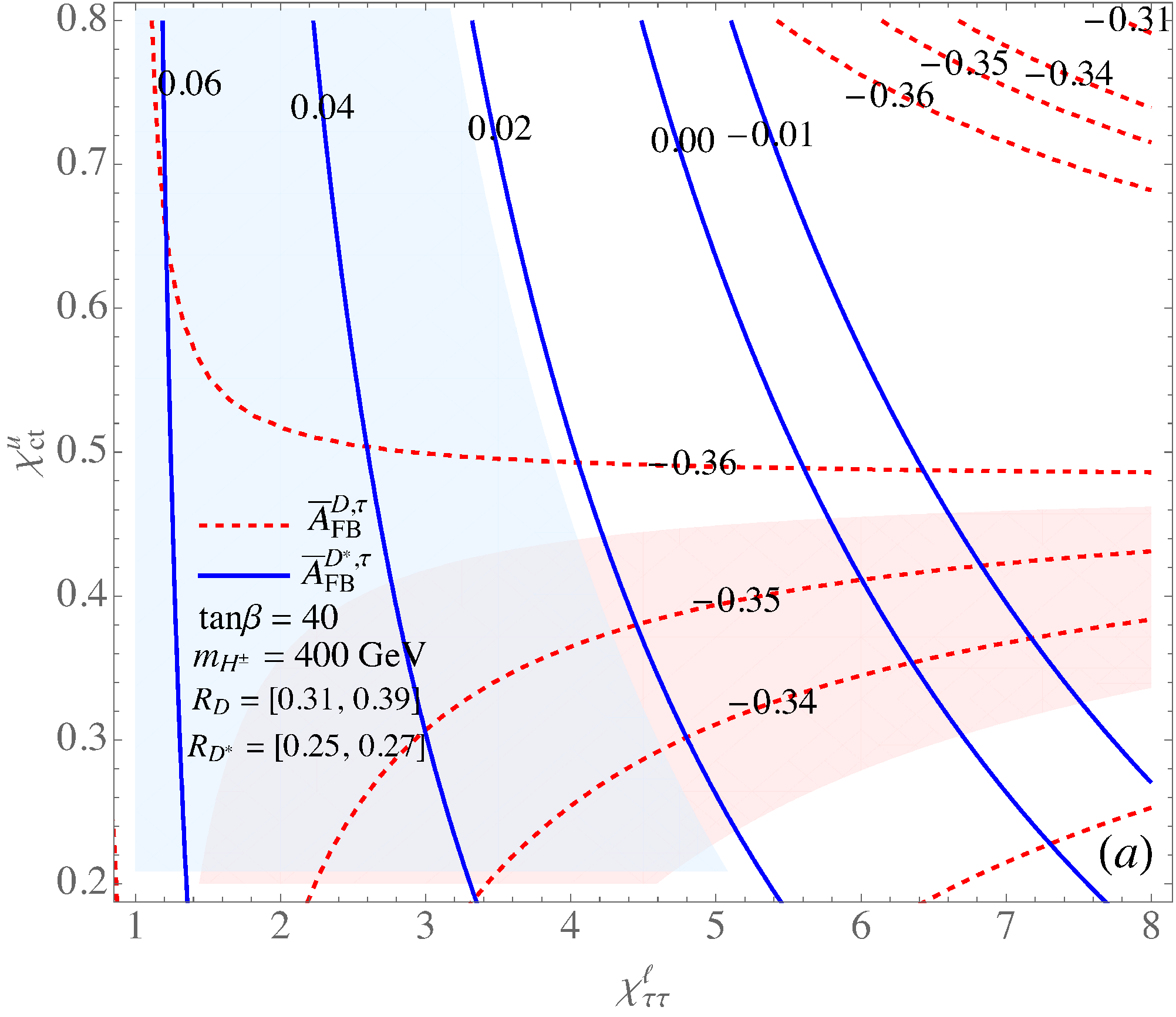}  
\includegraphics[width=75mm]{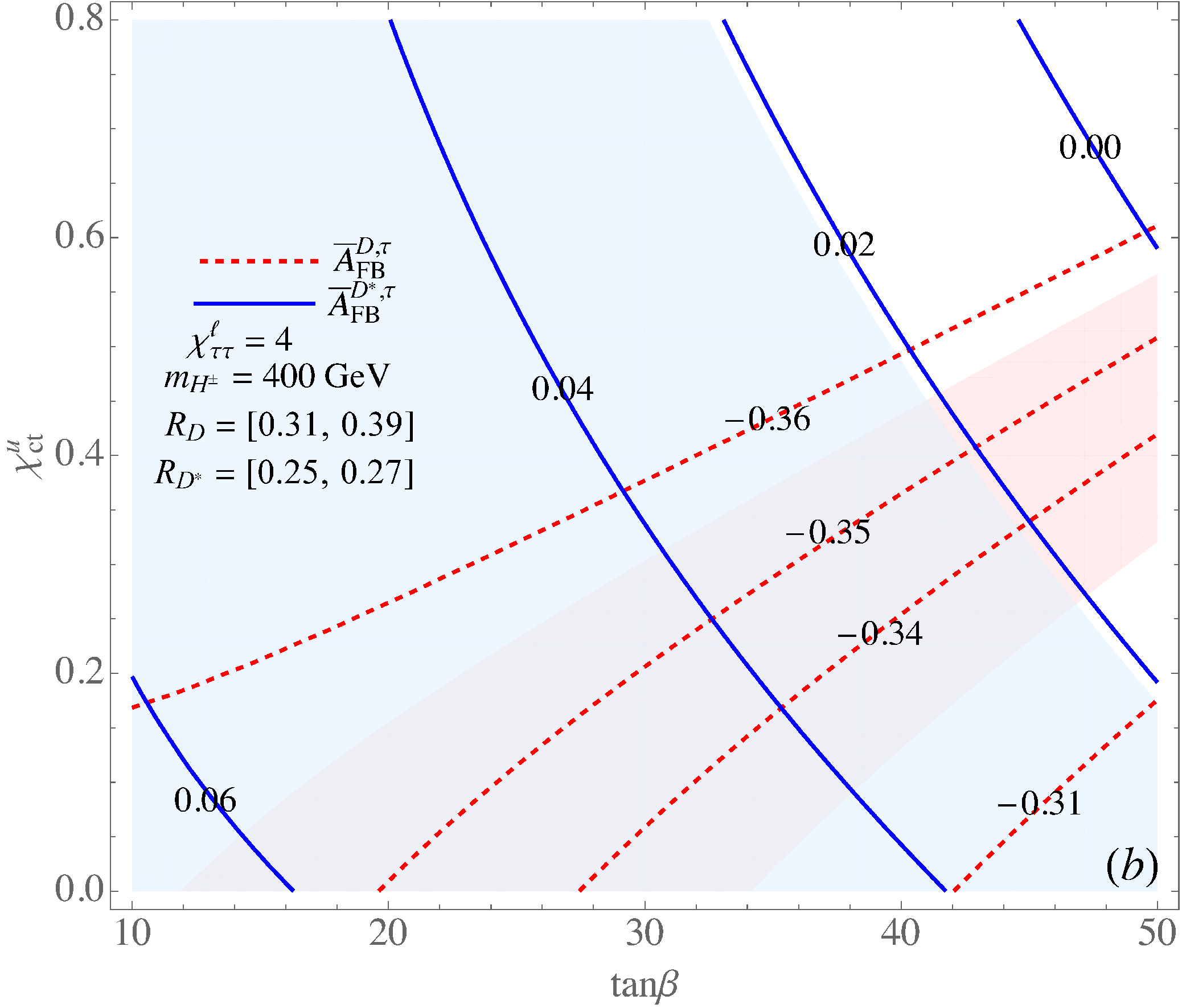} 
\caption{The legend is the same as that in Fig.~\ref{fig:Ptau}, but for FBA, defined in Eq.~(\ref{eq:Ifba}).}
\label{fig:AFB}
\end{figure}

\begin{table}[htp]
\caption{Values of integrated FBA, $R_{D(D^*)}$, and $P^\tau_{D(D^*)}$ with and without charged-Higgs effects in some chosen benchmarks of $(\chi^u_{ct}, \chi^\ell_{\ell \ell})$, where we have fixed $m_{H^\pm}=400$ GeV and $\tan\beta=40$. } \label{tab:FBA}
\begin{ruledtabular}
\begin{tabular}{c|ccccc}
$(\chi^u_{ct},\, \chi^\ell_{\tau\tau})$ &  $(0, 0)$ &  $( 0.2,\, 3 )$ & $( 0.2,\, 4 )$  &   $(0.3,\, 3)$  & $( 0.3,\, 4)$ \\ \hline
$\bar A^{D,\tau}_{FB}$  &  $-0.359$ &  $-0.344$ & $-0.335$  &  $-0.350$ & $-0.344$ \\ \hline
$\bar A^{D^*, \tau}_{FB}$ &  0.064 & $0.043$ & $0.033$ & $0.040$ & $0.029$  \\ \hline 
$R_D$ & 0.306 & 0.364 & 0.396 & 0.343 & 0.362 \\ \hline 
$R_{D^*}$ & 0.257 & 0.264 & 0.268 & 0.266 & 0.270  \\ \hline
$P^\tau_{D}$ & 0.324 & 0.432 & 0.479 & 0.396 & 0.429 \\ \hline
$P^\tau_{D^*}$ & $-0.500$ & $-0.459$ & $-0.438$ & $-0.453$ & $-0.428$ 
\end{tabular}
\end{ruledtabular}

\end{table}%

\section{Summary}

We studied the charged-Higgs $H^\pm$ effects on the $\bar B \to ( D, D^*) \ell \bar \nu_{\ell}$ decays  in a generic two-Higgs-doublet model. In order to parametrize the new $H^\pm$ Yukawa couplings to the quarks and leptons, we employ the Cheng-Sher ansatz. Accordingly, the third-generation $b$-quark and $\tau$-lepton related processes are dominant and can then be enhanced using a scheme with large $\tan\beta$.   Based on this study, it can be seen that two parameters ( $\chi^u_{ct}$ \& $\chi^\ell_{\tau \tau}$)  are required to  explain the $R_{D}$ and $R_{D^*}$ excesses. However, when the constraint from the $B_c \to \tau \bar \nu_\tau$ decay with $BR(B_c\to \tau \bar \nu_\tau) < 30\%$ is applied, $R_{D}$ can be still significantly enhanced while $R_{D^*}$ can only have a slight change.  The $\tau$ polarizations in the $\bar B \to (D, D^*) \tau \bar\nu_\tau$ decays were calculated, and  it was found that they are sensitive to the $H^\pm$ effects. The integrated $\tau$-lepton forward-backward asymmetries were studied. We found that the asymmetry of $\bar B \to D^* \tau \bar \nu_\tau$ is more sensitive to the $H^\pm$ effects. Although the sign of $\bar A^{D^*,\tau}_{FB}$   can be reversed in some parameter space,  when the constraint from $B_c\to \tau \bar\nu_\tau$  is included, the sign  is only positive.

}

\section*{Acknowledgments}

This work was partially supported by the Ministry of Science and Technology of Taiwan,  under grant MOST-103-2112-M-006-004-MY3 (CHC).

\appendix

\section*{Appendix}

In order to derive the charged lepton helicity amplitudes in the $\bar B\to M \ell \bar \nu_\ell$ decay, we need the specific spinor states of a charged lepton and  neutrino in the $q^2$ rest frame. Let $p=(E, \vec{p})$ be the four-momentum of a spin-1/2 particle, the solutions of the Dirac equation for positive and negative energy are expressed as:
 \begin{align}
 u_{\pm} (p) &= \frac{1}{\sqrt{E+m} }
\left(
\begin{array}{c}
 \sqrt{E+m} \chi_{\pm} ( \vec{p})    \\
    \vec{\sigma}\cdot \vec{p} \chi_{\pm}(\vec{p})  
\end{array}
\right)\,, \quad  %
v_{\pm} (p) &= \frac{1}{\sqrt{E+m} }
\left(
\begin{array}{c}
  \vec{\sigma}\cdot \vec{p} \chi_{\mp}(\vec{p})    \\
     \sqrt{E+m} \chi_{\mp} ( \vec{p}) 
\end{array}
\right)\,,
 \end{align}
 where the $\pm$ indices in $\chi$ are the eigenvalues of $\vec{\sigma}\cdot \vec{p} /|\vec{p}|$, and  $+/-$ denote the left-/right-handed states, respectively. 
If the spatial momentum of a particle is taken as $\vec{p}= p(\sin\theta \cos\phi, \sin\theta\sin\phi, \cos\theta)$, the eigenstates of $\vec{\sigma} \cdot \vec{p}$ can be found as:
 \begin{align}
 \chi_+ (\vec{p}) = \left(
\begin{array}{c}
\cos\frac{\theta}{2}    \\
   e^{i\phi}
 \sin\frac{\theta}{2}\end{array}
\right)\,, \quad   \chi_{-}(\vec{p})= \left(
\begin{array}{c}
\sin\frac{\theta}{2}    \\
  - e^{i\phi} \cos\frac{\theta}{2}\end{array}
\right)\,.
 \end{align}
 With the Pauli-Dirac representation of  $\gamma$-matrices, which are defined as:
 \begin{align}
 \gamma^0 = \left(
\begin{array}{cc}
{\bf 1} & 0    \\
   0 &  -{\bf 1} \end{array}
\right)\,, 
\quad \gamma^i =  \left(
\begin{array}{cc}
0 &  \sigma^i   \\
   -\sigma^i & 0\end{array}
\right)\,,  \quad
 \gamma_5=\gamma^5 = \left(
\begin{array}{cc}
0 & {\bf 1}    \\
   {\bf 1} &  0 \end{array}
\right)\,,
 \end{align}
   we get $\bar \ell_{u_\pm} [...] (1-\gamma_5) \nu_{v_+} = 2 \bar \ell_{u_\pm} [...] \nu_{v_{+}}$ and $\bar \ell [...] (1-\gamma_5) \nu_{v_+}=0$ when  $m_\nu=0$ is applied, in which $[...] =\{ 1, \gamma^\mu, \sigma^{\mu \nu} \}$.  
 
 For simplifying the derivations of $\bar\ell_{u_{\pm}} [...] (1-\gamma_5) \nu_{v_+}$, we define some useful  polarization vectors as: 
 \begin{align}
 |\vec{P}| \epsilon^\mu_X & \equiv  P^\mu - \frac{P\cdot q }{q^2} q^\mu\,, \quad \epsilon^\mu_X \epsilon_{X \mu} =-1 \,, \nonumber \\
 \frac{E_{D^*}}{m_{D^*}} \epsilon^\mu_Z & \equiv \epsilon^\mu (L) - \frac{\epsilon \cdot q }{q^2} q^\mu \,, 
 \quad \sqrt{\frac{\lambda_{D^*}}{2}} e^\mu_{D^*}(T)  \equiv  \varepsilon^{\mu \nu \rho \sigma} \epsilon_\nu(T) P_\rho q_\sigma
 \end{align}
 with $|\vec{P}|= \sqrt{\lambda_M}/\sqrt{q^2}$. 
 According to the chosen coordinates in the $q^2$ rest frame, the leptonic current associated with a specific charged lepton helicity can be derived as follows:  for the $B\to D$ case, we get: 
 \begin{align}
 \bar\ell_{u_+} \slashed{\epsilon}_X (1- \gamma_5)  \nu_{v_+} &= 2 m_\ell \beta_\ell \cos\theta_\ell\,, \nonumber \\
 \bar\ell_{u_+}  (1- \gamma_5)  \nu_{v_+} &= -2 \sqrt{q^2} \beta_\ell  \,, \nonumber  \\
 \bar\ell_{u_-} \slashed{\epsilon}_X (1- \gamma_5)  \nu_{v_+} &= -2 \sqrt{q^2} \beta_\ell \sin\theta_\ell\,, \nonumber \\
 \bar\ell_{u_-}  (1- \gamma_5)  \nu_{v_+} &= 0
 \end{align}
 with $\beta_\ell = \sqrt{1 - m^2_\ell/q^2}$.  For the $B\to D^*$ case, the $D^*$ longitudinal parts are obtained as:
 %
 %

%
  %
%
%
\begin{align}
  \bar\ell_{u_+} \slashed{\epsilon}_Z (1- \gamma_5)  \nu_{v_+} &= 2m_\ell \beta_\ell  \cos\theta_\ell \,, \\
    \bar\ell_{u_-} \slashed{\epsilon}_Z (1- \gamma_5)  \nu_{v_+} &= - 2\sqrt{q^2} \beta_\ell \sin\theta_\ell\,, 
\end{align}
while the two $D^*$ transverse parts are respectively given as:
\begin{align}
\bar\ell_{u_+} \slashed{e}_{D^*}(T) (1- \gamma_5)  \nu_{v_+} &=  -2 m_\ell \beta_\ell  \left\{
\begin{array}{c}
 \frac{i}{\sqrt{2}} \sin\theta_\ell  e^{-i \phi}  ~(T=+) \,,   \\
   \frac{i }{\sqrt{2}} \sin\theta_\ell  e^{i \phi}  ~ (T= -) \,,
  \end{array}
   \right.
\end{align}

\begin{align}
\bar\ell_{u_-} \slashed{e}_{D^*}(T) (1- \gamma_5)  \nu_{v_+} &=  -2\sqrt{q^2} \beta_\ell  \left\{
\begin{array}{c}
 \frac{- i }{\sqrt{2}} (1- \cos\theta_\ell)  e^{-i \phi}  ~(T=+) \,,   \\
   \frac{i }{\sqrt{2}} (1 + \cos\theta_\ell ) e^{i \phi}  ~ (T= -) \,.
  \end{array}
   \right.\,, 
\end{align}

The differential decay rates shown in Eq.~(\ref{eq:ang_Ga}) are functions of $q^2$ and $\theta_{\ell}$. After integrating out 
the polar angle, the differential decay rate with each lepton helicity as a function of $q^2$  can be obtained as follows: For the $\bar B \to D \ell \bar \nu_\ell$ decay,  they can be shown as:
 \begin{align}
 \frac{d \Gamma^{h=\pm}_{D}}{dq^2 } & = \frac{G^2_F |V_{cb}|^2 \sqrt{\lambda_D} \beta^4_\ell}{256 \pi^3 m^3_B} H^{\pm}_{D}\,, \label{eq:Gah_D}\\
 H^+_{D} &= \frac{2 m^2_\ell }{3 q^2 }\lambda_D  F^2_1 + 2 m^2_\ell q^2 |X^0_D|^2\,, ~
 H^-_{D}  =\frac{4}{3} \lambda_D F^2_1 \,;
 \end{align}
and for the $\bar B\to D^* \ell \nu_\ell$ decay, they are expressed as:
  \begin{align}
 \frac{d \Gamma^{\lambda, h=\pm}_{D^*}}{dq^2 } & = \frac{G^2_F |V_{cb}|^2 \sqrt{\lambda_{D^*}} \beta^4_\ell}{256 \pi^3 m^3_B} H^{\lambda, \pm}_{D^*}\,, \label{eq:Gah_D*} \\
 H^{L,+}_{D^*} &= \frac{2 m^2_\ell  }{3 } |h^0_{D^*}|^2 +  \frac{2 m^2_\ell }{q^2} \lambda_{D^*} |X^0_{D^*}|^2\,, ~ H^{L,-}_{D^*}  =\frac{4 q^2 }{3}  |h^0_{D^*}|^2\,, \\
 H^{T=\pm,+}_{D^*} &= \frac{2 m^2_\ell  }{3 } |h^{\pm}_{D^*}|^2 \,, ~ H^{T=\pm,-}_{D^*} = \frac{4 q^2 }{3 } |h^{\pm}_{D^*}|^2
 \end{align}


\end{document}